\documentclass{article}
\usepackage{multicol}
\usepackage{amsfonts, amsmath, amsthm, graphicx}
\usepackage{subcaption}
\usepackage{comment}

\theoremstyle{plain}
\newtheorem{teo}{Theorem}
\newtheorem{claim}{Claim}
\newtheorem{case}{Case}
\newtheorem{cor}[teo]{Corollary}

\theoremstyle{remark}
\newtheorem{remark}{\textbf{\textit{Remark}}}

\begin{document}

\title{On nested and 2-nested graphs: two subclasses of graphs between threshold and split graphs.}
\author{Nina Pardal \and Guillermo A. Dur\'an \and Luciano N. Grippo \and Mart\'in D. Safe \thanks
{Supported by PIO CONICET UNGS-144-20140100011-CO, UNS Grants PGI 24/L103 and PGI 24/L115, ANPCyT PICT-2017-1315, and MATH-AmSud 18-MATH-01. This work was partially financed by ISCI, Chile (CONICYT PIA FB0816; ICM P-05-004-F), ANPCyT PICT grant 2015-2218 (Argentina) and UBACyT grant 20020170100495BA (Argentina). 
\newline
\emph{2000 AMS Subject Classification:} 05C75, 05C50 and 05C38. \newline
\emph{Keywords:} circle graphs, split graphs, 2-coloring, consecutive ones property, characterization}}
\date{}
\maketitle

\begin{abstract}
A $(0,1)$-matrix has the Consecutive Ones Property (C1P) for the rows if there is a permutation of its columns such that the ones in each row appear consecutively. We say a $(0, 1)$-matrix is nested if it has the consecutive ones property for the rows (C1P) and every two rows are either disjoint or nested. We say a $(0, 1)$-matrix is 2-nested if it has the C1P and admits a partition of its rows into two sets such that the submatrix induced by each of these sets is nested. We say a split graph $G$ with split partition $(K, S)$ is nested (resp.\ 2-nested) if the matrix $A(S, K)$ which indicates the adjacency between vertices in $S$ and $K$ is nested (resp.\ 2-nested). In this work, we characterize nested and 2-nested matrices by minimal forbidden submatrices. This characterization leads to a minimal forbidden induced subgraph characterization for these classes of graphs, which are a superclass of threshold graphs and a subclass of split and circle graphs. 
\end{abstract}

\section{Introduction} 

Let $A=(a_{ij})$ be a $n\times m$ $(0,1)$-matrix.
We denote $a_{i.}$ and $a_{.j}$ the $i$th row and the $j$th column of matrix $A$.
Let $l_i = \min\{ j \colon\,a_{ij} = 1 \}$ and $r_i = \max\{ j \colon\,a_{ij} = 1 \}$ for each $i\in\{1,\ldots,n\}$. Two rows \emph{$a_{i.}$ and $a_{k.}$ are disjoint} if there is no $j$ such that $a_{ij} = a_{kj} = 1$.
We say that $a_{i.}$ is \emph{contained} in $a_{k.}$ if for each $j$ such that $a_{ij} = 1$ also $a_{kj} = 1$. We say that \emph{$a_{i.}$ and $a_{k.}$ are nested} if $a_{i.}$ is contained in $a_{k.}$ or $a_{k.}$ is contained in $a_{i.}$.
Finally, we say that \emph{$a_{i.}$ and $a_{k.}$ start} (resp.\ \emph{end}) \emph{in the same column} if $l_i = l_k$ (resp.\ $r_i = r_k$), 
and we say \emph{$a_{i.}$ and $a_{k.}$ start (end) in different columns} otherwise. 
We say a $(0,1)$-matrix $A$ has the \emph{consecutive ones property for the rows }(for short, C1P) if there is permutation of the columns of $A$ such that the 1's in each row appear consecutively. 
 Tucker characterized all the minimal forbidden submatrices for the C1P, later known as \emph{Tucker matrices}. For the complete list of Tucker matrices, see \cite{T72}, where a graphic representation of them can be found in Figure 3. 
 
We say a $(0,1)$-matrix is \emph{nested} if it has the consecutive ones property for the rows (C1P) and every two rows are either disjoint or nested. We say a $(0,1)$-matrix is \emph{2-nested} if it has the C1P for the rows and there is a partition $S_1, S_2$ of the rows such that each submatrix obtained is nested.

All graphs in this work are simple, undirected, with no loops and no multiple edges. The pair $(K,S)$ is a \emph{split partition} of a graph $G$ if $\{K,S\}$ is a partition of the vertex set of $G$ and the vertices of $K$ (resp.\ $S$) are pairwise adjacent (resp.\ nonadjacent), and we denote it $G=(K,S)$. A graph $G$ is a \emph{split graph} if it admits some split partition. Let $G$ be a split graph with split partition $(K,S)$, $n=\vert S\vert$, and $m=\vert K\vert$. Let $s_1, \ldots, s_n$ and $v_1, \ldots, v_m$ be linear orderings of $S$ and $K$, respectively. Let $A= A(S,K)$ be the $n\times m$ matrix defined by $A(i,j)=1$ if $s_i$ is adjacent to $v_j$ and $A(i,j)=0$, otherwise.

A split graph $G = (K,S)$ is \emph{nested} (resp.\ \emph{2-nested}) if there is a linear ordering $\Pi$ of $K$, such that the associated matrix $A(S,K)$ is nested (resp.\ 2-nested) and if its columns are ordered as in $\Pi$ then the ones in each row occur in consecutive columns.

Circle graphs \cite{EI71} are intersection graphs of chords in
a circle. These graphs were characterized by Bouchet \cite{B94} in 1994 by forbidden induced subgraphs under local complementation. However, no complete characterizations of circle graphs by forbidden induced subgraphs of the graph itself are known. It follows from the definition that nested and 2-nested graphs are common subclasses of circle graphs. 
Furthermore, nested and 2-nested graphs are also a superclass of threshold graphs (see Golumbic \cite{G04} for more details on these definitions).

The problem of characterizing 2-nested graphs by minimal forbidden induced subgraphs arises as a natural subproblem in our ongoing efforts to obtain the same kind of characterization of those split graphs that are circle graphs. 
We started by considering a split graph $H$ such that $H$ is minimally non-circle. Since comparability graphs are a subclass of circle graphs, in particular $H$ is not a comparability graph. Notice that permutation graphs are those comparability graphs for which their complement is also a comparability graph. It is easy to prove that permutation graphs are precisely those circle graphs having a circle model with an equator. See Gallai \cite{G67} for the complete list of minimal forbidden subgraphs of comparability graphs. Using the list of minimal forbidden subgraphs of comparability graphs and the fact that $H$ is also a split graph, we conclude that $H$ contains either a tent, a 4-tent, a co-4-tent or a net as a subgraph. We first considered the case in which $H$ contains an induced tent as a subgraph, thus reaching a problem when trying to give a circle model for $H$.
Once analyzed the compatibilities between the vertices in the complete and independent partitions of such a graph, it arises that there is exactly one subclass --which we denoted $\alpha$-- of independent vertices for which both endpoints of each vertex could be entirely drawn in two distinct areas of the circle model, when for every other vertex there is a unique possible placement. Hence, for the subgraph induced by taking the tent graph union the subclass $\alpha$ to admit a circle model, the subclass $\alpha$ must be partitioned into two disjoint subsets such that, for each subset, every pair of vertices are either disjoint or nested, thus leading to the definition of 2-nested graphs.

\vspace{-3mm}

\section{Nested matrices}

\begin{figure}[h!]
	\centering  
	\includegraphics[scale=.37]{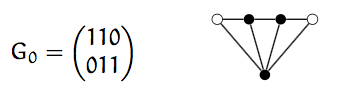}
	\vspace{-2mm}
	\caption{The $G_0$ matrix and the gem graph} \label{fig:forb_nested}
\end{figure}

We begin by giving the following characterization of nested matrices. 
\begin{teo}\label{thm:teo-nested-matrix}
	A $(0,1)$-matrix is nested if and only if it contains no $G_0$ as a submatrix (see Figure~\ref{fig:forb_nested}).
\end{teo}
\vspace{-3mm}

\vspace{-3mm}
\begin{proof}
Since no Tucker matrix has the C1P and the rows of $G_0$ are neither disjoint nor nested, no nested matrix contains a Tucker matrix or $G_0$ as submatrices. Conversely, as each Tucker matrix contains $G_0$ as a submatrix, every matrix containing no $G_0$ as a submatrix is a nested matrix.
\end{proof}

\begin{cor}
	A split graph is nested if and only if it contains no induced gem.
\end{cor}

\vspace{-4mm}
\section{2-nested matrices}

\vspace{-1mm}
We define the following matrices, since they play an important role in the sequel. 

\vspace{-2mm}

\begin{figure}[h!]
	\centering
	\begin{align*}
			F_0= \left(\begin{smallmatrix}
				11100\\
				01110\\
				00111\\
			\end{smallmatrix}\right)
			&&
			F_1(k)= \left(\begin{smallmatrix}
				011...111\\
				111...110\\
				000...011\\
				000...110\\
				.   .   .   .   . \\
				.   .   .   .   . \\
				.   .   .   .   . \\
				110...000\\
			\end{smallmatrix}\right)
			&&
			F_2(k)= \left(\begin{smallmatrix}
				0111...10\\
				1100...00\\
				0110...00\\
				.   .   .   .   . \\
				.   .   .   .   . \\
				.   .   .   .   . \\
				0000...11\\
			\end{smallmatrix}\right)
	\end{align*}
	\vspace{-3mm}
	\caption{$F_0$, $F_1(k) \in \{0,1\}^{k \times k-1}$, and $F_2(k) \in \{0,1\}^{k \times k}$, for any odd $k \geq 5$.}
\end{figure} \label{fig:forb_B-A}

\vspace{-4mm}
\begin{teo}
	A $(0,1)$-matrix $A$ is 2-nested if and only if there is a linear ordering $\Pi$ of the columns such that the matrix $A$ with its columns ordered according to $\Pi$ does not contain any Tucker matrix, or $F_0$, $F_1(k)$, $F_2(k)$ for every odd $k \geq 5$ as a configuration.
\end{teo}

\vspace{-1.5mm}

We define the auxiliary graph $H(A)=(V,E)$ where the vertex set $V= \{ w_1, \ldots, w_n \}$ has one vertex for each row in $A$, 
and two vertices $w_i$ and $w_k$ in $V$ are adjacent if and only if the rows $a_{i.}$ and $a_{k.}$ are neither disjoint nor nested. By abuse of language, $w_i$ will refer to both the vertex $w_i$ in $H(A)$ and the row $a_{i.}$ of $A$. In particular, the definitions given in the introduction apply to the vertices in $H(A)$; i.e., we say two vertices $w_i$ and $w_k$ in $H(A)$ are \emph{nested} (resp.\ \emph{disjoint}) if the corresponding rows $a_{i.}$ and $a_{k.}$ are nested (resp.\ disjoint). And two vertices $w_i$ and $w_k$ in $H(A)$ \emph{start} (resp.\ \emph{end}) \emph{in the same column} if the corresponding rows $a_{i.}$ and $a_{k.}$ start (resp.\ end) in the same column. 
It follows from the definition of 2-nested matrices that $A$ is a 2-nested matrix if and only if there is a bicoloring of the auxiliary graph $H(A)$ or, equivalently, if $H(A)$ is bipartite (i.e., $H(A)$ does not contain cycles of odd length).

\begin{proof} 
	Since $A$ admits a C1P, then $A$ contains no Tucker matrices. Moreover, if $A$ contains $F_0$, $F_1(k)$ or $F_2(k)$ for some odd $k \geq 5$, since the corresponding subgraphs in $H(A)$ of every such matrix induces an odd cycle, then it follows that $H(A)$ does not admit a proper 2-coloring and this results in a contradiction. Therefore, $A$ does not contain any $F_0$, $F_1(k)$ or $F_2(k)$ for any odd $k\geq 5$ as a configuration.	

	Conversely, let $\Pi$ be a linear ordering of the columns such that the matrix $A$ does not contain any $F_0, F_1(k), F_2(k)$ for any odd $k\geq 5$ or Tucker matrices as configurations.
	Due to Tucker's Theorem, since there are no Tucker submatrices in $A$,
	the matrix $A$ has the C1P. 
	
	Towards a contradiction, suppose that the auxiliary graph $H(A)$ is not bipartite. Hence there is an induced odd cycle $C$ in $H(A)$.

	Suppose first that $H(A)$ has an induced odd cycle $C = w_1, w_2, w_3, w_1 $ of length 3, and suppose without loss of generality that the first rows of $A$ are those corresponding to the cycle $C$. 
	Since $w_1$ and $w_2$ are adjacent, both begin and end in different columns. The same holds for $w_2$ and $w_3$, and $w_1$ and $w_3$. We assume without loss of generality that the vertices start in the order of the cycle, in other words, that $l_1 < l_2 < l_3$. 
	
	Since $w_1$ starts first, it is clear that $a_{2 l_1} = a_{3 l_1} = 0$, thus the column $a_{. l_1}$ of $A$ is the same as the first column of the matrix $F_0$.

	Since $A$ has the C1P and $w_1$ and $w_2$ are adjacent, then $a_{1 l_2} = 1$. 
	As stated before, $w_2$ starts before $w_3$ and thus $a_{3 l_2} = 0$. Hence, column $a_{. l_2}$ is equal to the second column of $F_0$.
	
	The third column of $F_0$ is $a_{. l_3}$, for $w_3$ is adjacent to $w_1$ and $w_2$, hence it is straightforward that $a_{1 l_3} = a_{2 l_3} = a_{3 l_3} = 1$.
	
	To find the next column of $F_0$, let us look at column $a_{. (r_1+1)}$. Notice that $r_1 + 1 > l_3$. Since $w_1$ is adjacent to $w_2$ and $w_3$, and $w_2$ and $w_3$ both start after $w_1$, then necessarily $a_{2 (r_1+1)} = a_{3 (r_1+1)} = 1$, and thus $a_{. (r_1+1)}$ is equal to the fourth column of $F_0$.
	
	Finally, we look at the column $a_{. (r_2 +1)}$. Notice that $r_2 +1 > r_1 + 1$. 
	Since $A$ has the C1P, $a_{1 (r_2 +1)} = 0$ and $r_2 +1 > r_1+1$, then $a_{1 (r_2+1)} =0$ and $a_{3 (r_2+1)} = 1$, which is equal to last column of $F_0$. Therefore we reached a contradiction that came from assuming that there is a cycle of length 3 in $H(A)$.
	
	Suppose now that $H(A)$ has an induced odd cycle $C = w_1, \ldots, w_k, w_1$ of length $k \geq 5$. We assume without loss of generality that the first $k$ rows of $A$ are those in $C$ and that $A$ is ordered according to the C1P.
	
	\begin{remark} \label{rem:2N_1}
		Let $w_i, w_j$ be vertices in $H(A)$. If $w_i$ and $w_j$ are adjacent and $w_i$ starts before $w_j$, then $a_{i r_i} = a_{j r_i} = 1$ and $a_{i (r_i+1)} = 0$, $a_{j (r_i +1)} = 1$.
	\end{remark}

\begin{remark} \label{rem:2N_2}
	If $l_{i-1} > l_i$ and $l_{i+1} > l_i$ for some $i = 3, \ldots, k-1$, then for all $j\geq i+1$, $w_j$ is nested in $w_{i-1}$. The same holds if $l_{i-1} < l_i$ and $l_{i+1} < l_i$.
	Since $l_{i-1} > l_i$ and $l_{i+1} > l_i$, then $w_{i-1}$ and $w_{i+1}$ are not disjoint, thus necessarily $w_{i+1}$ is nested in $w_{i-1}$. It follows from this argument that this holds for $j \geq i+1$.
\end{remark}

Notice that $w_2$ and $w_k$ are nonadjacent, hence they are either disjoint or nested. Using this fact and Remark \ref{rem:2N_1}, we split the proof into two cases.

\begin{case}
	$w_2$ and $w_k$ are nested 
\end{case}	
\vspace{-1mm}
We may assume without loss of generality that $w_k$ is nested in $w_2$, for if not, we can rearrange the cycle backwards as $w_1$, $w_k$, $w_{k-1}, \ldots, w_2, w_1$. Moreover, we will assume without loss of generality that both $w_2$ and $w_k$ start before $w_1$. First, we need the following Claim.

\begin{claim} \label{claim:2N_1-1}
	If $w_2$ and $w_k$ are nested, then $w_i$ is nested in $w_2$, for $i = 4, \ldots, k-1$.
\end{claim}

Suppose first that $w_1$ and $w_3$ are disjoint, and towards a contradiction suppose that $w_2$ and $w_4$ are disjoint. In this case, $l_4 < l_3 < r_4 < l_2 < r_3 < r_2$. 
The contradiction is clear if $k=5$.
If instead $k>5$ and $w_5$ starts before $w_4$, then $r_i < l_3$ for all $i > 5$, which contradicts the assumption that $w_k$ is nested in $w_2$. Hence, necessarily $w_5$ is nested in $w_3$ and $w_5$ and $w_2$ are disjoint. This implies that $l_3 < l_5 < r_4 < r_5 < l_2$ and once more, $r_i < l_2$ for all $i >5$, which contradicts the fact that $w_k$ is nested in $w_2$.

Suppose now that $w_3$ is nested in $w_1$. Towards a contradiction, suppose that $w_4$ is not nested in $w_2$. Thus, $w_2$ and $w_4$ are disjoint since they are nonadjacent vertices in $H(A)$.
Notice that, if $w_3$ is nested in $w_1$, then $l_2 < l_3$ and $r_2 < r_3$.
Furthermore, since $w_4$ is adjacent to $w_3$ and nonadjacent to $w_2$, then $l_3 < r_2 < l_4 < r_3 < r_4$. This holds for every odd $k\geq 5$.

If $k=5$, since $w_5$ is nested in $w_2$, then $r_5 < r_2 < l_4$, which results in a contradiction for $w_4$ and $w_5$ are adjacent.

Suppose that $k>5$. If $w_2$ and $w_i$ are disjoint for all $i= 5, \ldots, k-1$, then $w_{k-1}$ and $w_k$ are nonadjacent for $w_k$ is nested in $w_2$, which results in a contradiction. Conversely, if $w_i$ and $w_2$ are not disjoint for some $i > 3$, then they are adjacent, which also results in a contradiction that came from assuming that $w_2$ and $w_4$ are disjoint. Therefore, since $w_4$ is nested in $w_2$, $w_2$ and $w_i$ are nonadjacent and $w_i$ is adjacent to $w_{i+1}$ for all $i >4$, then necessarily $w_i$ is nested in $w_2$, which finishes the proof of the Claim. 

\begin{claim} \label{claim:2N_1-2}
	Suppose that $w_2$ and $w_k$ are nested. Then, if $w_3$ is nested in $w_1$, then $l_i > l_{i+1}$ for all $i=3, \ldots, k-1$. If instead $w_1$ and $w_3$ are disjoint, then $l_i < l_{i+1}$ for all $i=3, \ldots, k-1$.
\end{claim}

\vspace{-0.5mm}
Recall that, by the previous Claim, since $w_i$ is nested in $w_2$ for all $i = 4, \ldots, k $, in particular $w_4$ is nested in $w_2$. Moreover, since $w_3$ and $w_4$ are adjacent, notice that, if $w_3$ is nested in $w_1$, then $l_3 > l_4$, and if $w_1$ and $w_3$ are disjoint, then $l_3 < l_4$.

It follows from Remark \ref{rem:2N_2} that, if $l_5 > l_4$, then $w_i$ is nested in $w_3$ for all $i = 5, \ldots, k $, which contradicts the fact that $w_1$ and $w_{k-1}$ are adjacent. The proof of the first statement follows from applying this argument successively.

The second statement is proven analogously by applying Remark \ref{rem:2N_2} if $l_5 < l_4$, and afterwards successively for all $i > 4$. 

If $w_1$ and $w_3$ are disjoint, then we obtain $F_2(k)$ first, by putting the first row as the last row, and considering the submatrix given by columns $j_1 = l_1 -1$, $j_2 = l_3$, $\ldots$, $j_i = l_{i+1}$, $\ldots$, $j_k = r_1+1$ (using the new ordering of the rows).
If instead $w_3$ is nested in $w_1$, then we obtain $F_1(k)$ by taking the submatrix given by the columns $j_1 = l_1 -1$, $j_2 = r_k$, $\ldots$, $j_i = l_{k -i +2}$, $\ldots$, $j_{k-1} = r_3$.

\begin{case}
	$w_2$ and $w_k$ are disjoint 
\end{case}	
\vspace{-1mm}
We assume without loss of generality that $l_2 < l_1$ and $l_k > l_1$.

\begin{claim} \label{claim:2N_2-1}
	If $w_2$ and $w_k$ are disjoint, then $l_i < l_{i+1}$ for all $i=2, \ldots, k-1$.
\end{claim}

Notice first that, in this case, $w_i$ is nested in $w_1$, for all $i= 3, \ldots, k-1$. If not, then using Remark \ref{rem:2N_2}, we notice that it is not possible for the vertices $w_1, \ldots, w_k$ to induce a cycle. This implies, in particular, that $w_3$ is nested in $w_1$ and thus $l_2 < l_3$.
Furthermore, using this and the same remark, we conclude that $l_i < l_{i+1}$ for all $i = 2, \ldots, k-1$, therefore proving Claim \ref{claim:2N_2-1}. 
 
In this case, we obtain $F_2(k)$ by considering the submatrix given by the columns $j_1 = l_1 -1$, $j_2 = l_3$, $\ldots$, $j_i = l_{i+1}$, $\ldots$, $j_k = r_1+1$.\qedhere

\end{proof}

\begin{tabular}{p{0.5\textwidth}p{0.5\textwidth}}
{\parindent=0pt
\obeylines
Nina Pardal
Instituto de C\'alculo
Universidad de Buenos Aires
Buenos Aires, Argentina
npardal@ic.fcen.uba.ar
\par
}
&
{\parindent=0pt
\obeylines
Guillermo A. Dur\'an
Instituto de C\'alculo
Universidad de Buenos Aires
Buenos Aires, Argentina
gduran@ic.fcen.uba.ar
\par
}\\[1em]
{\parindent=0pt
\obeylines
Luciano N. Grippo 
Instituto de Ciencias
UNGS
Buenos Aires, Argentina
lgrippo@ungs.edu.ar
\par
}

&
{\parindent=0pt
\obeylines
Mart\'in D. Safe
INMABB,  Depto. de Matem\'atica
Univ. Nacional del Sur--CONICET
Bah\'ia Blanca, Argentina
msafe@uns.edu.ar
\par
}
\end{tabular}

\end{document}